\documentclass[aps,11pt,preprint,superscriptaddress,showkeys,floatfix]{revtex4-1}
\usepackage{multirow}
\usepackage{graphicx}
\usepackage{color}

\begin{document}

\title{Thermoelectric properties of the misfit cobaltate Ca$_3$Co$_4$O$_9$}

\author{Bin Amin}
\affiliation{KAUST, PSE Division, Thuwal 23955-6900, Kingdom of Saudi Arabia}

\author{Ulrich Eckern}
\email{ulrich.eckern@physik.uni-augsburg.de}
\affiliation{Universit\"at Augsburg, Institut f\"ur Physik, 86135 Augsburg, Germany}

\author{Udo Schwingenschl\"ogl}
\email{udo.schwingenschlogl@kaust.edu.sa}
\affiliation{KAUST, PSE Division, Thuwal 23955-6900, Kingdom of Saudi Arabia}

\keywords{misfit cobaltate, thermoelectricity, density functional theory, relaxation time approximation}

\begin{abstract}
The layered misfit cobaltate Ca$_3$Co$_4$O$_9$, also known as
Ca$_2$CoO$_3$[CoO$_2$]$_{1.62}$, is a promising p-type thermoelectric oxide. Employing density
functional theory, we study its electronic structure and determine, on the basis of Boltzmann
theory within the constant-relaxation-time approximation, the thermoelectric transport coefficients. 
The dependence on strain and temperature is determined. In particular, we find that the $xx$-component
of the thermopower is strongly enhanced, while the $yy$-component is strongly reduced,
when applying 2\% tensile strain. A similar anisotropy is also found in the power factor. The
temperature dependence of the conductivity in the $a$-$b$ plane is found to be rather weak above 200 K,
which clearly indicates that the experimentally observed transport properties are dominated by
inhomogeneities arising during sample growth, i.e., are not intrinsic.
\end{abstract}

\maketitle

\section{Introduction}
Since the discovery of a large thermoelectric power in the layered cobaltate NaCoO$_2$ two decades
ago \cite{Terasaki97}, this system and closely related ones, particularly Ca$_3$Co$_4$O$_9$ (CCO),
have been studied intensively, both
experimentally \cite{Masset00,Lambert01,Xu02,Shikano03,Takeuchi04a,Takeuchi04b,Hu05,Limelette05,Lee06,
Eng06,Limelette06,Wakisaka08,Muguerra08,Tyson09,Kenfaui10,Tang11,Huang11,Qiao11,Praso12,Wu12,Madre13,Wu13,Jood13,
Kraus14,Huang15} 
as well as theoretically \cite{Asahi02,Singh07a,Rebola12,Soret12,Saeed12,Lemal17}; 
see also Refs.~\onlinecite{Singh07b,Ohta08,He11} for recent reviews.
Layered cobalt oxides,  such as CCO, are of interest because of the
close proximity of the rock-salt Ca$_2$CoO$_3$ and hexagonal CoO$_2$ subsystems
which results in a high Seebeck coefficient, low thermal conductivity,
and, hence, high figure of merit at elevated temperature.

For example,
Wakisaka et al.\ \cite{Wakisaka08} have studied the electronic structure
using x-ray photoemission spectroscopy, ultraviolet photoemission spectroscopy, and
unrestricted Hartree-Fock calculations. They derive a charge-transfer energy of 1 eV between
the Co sites in the Ca$_2$CoO$_3$ and CoO$_2$ subsystems as well as a $d$-$d$ Coulomb interaction 
energy of 6.5 eV. Huang et al.\ \cite{Huang11} have investigated the structural, electrical,
and thermal transport properties of CCO ceramics sintered in a high
magnetic field, arguing that this technique is efficient for enhancing the thermopower.
High-resolution photoemission spectroscopy has been used by Takeuchi et al.\ \cite{Takeuchi04a,Takeuchi04b}
to investigate the electronic structure. These authors argue that the high
thermopower is due to the metallic conduction in the hexagonal
CoO$_2$ layers. According to Refs.~\onlinecite{Muguerra08,Lambert01}, the rock-salt
subsystem mainly shortens the phonon mean free path and thus lowers the thermal conductivity,
while the hexagonal subsystem is responsible for the electrical conductivity. In addition, the
magneto-transport properties are remarkable \cite{Masset00,Eng06,Tang11}; in particular, a
high-temperature (above 600 K) spin state transition has been discovered very
recently \cite{Altin14,Karaman16}.

First principles calculations have been performed by Asahi et al.\ \cite{Asahi02} using
a 3/2 ratio for the $b_1$ (rock-salt) and $b_2$ (CoO$_2$) lattice parameters, 
$(b_1/b_2)_{\rm exp} = 1.62$, arguing that the optimized
structure with distorted octahedra is in agreement with experiment. The authors report
on p-type conductivity in the rock-salt subsystem, where the Fermi energy
lies in the crystal-field gap of the $3d$ states of the CoO$_2$ subsystem.
Tyson et al.\ \cite{Tyson09} have combined temperature-dependent local
structure measurements with first-principles calculations to develop a three-dimensional 
structure model of misfit Ca$_3$Co$_4$O$_9$. They find a low coordination of Co in the 
rock-salt layer, which is due to the formation of Co chains in the $a$-$b$ plane linked to the 
CaO layers by Co-O bonds oriented along the $c$ axis, enabling high electrical and
low thermal conductivity.

Combining experiment, ab-initio calculations,
and semiclassical Boltzmann transport theory to investigate the anisotropy of the
thermopower in Ca$_3$Co$_4$O$_9$ single crystals, Tang et al.\ \cite{Tang11}
find a strongly anisotropic topology of the Fermi surface.
R\'ebola et al.\ \cite{Rebola12} have investigated the atomic and electronic
structures employing density functional theory (DFT) and its extension (DFT+$U$). 
Both methods lead to good agreement of the structural parameters with experiment.

In this work, we consider the 5/3 rational approximation, optimize the structure 
employing density functional theory (in the VASP implementation), and determine
the thermoelectric transport coefficients on the basis of BoltzTrap. The transport
properties are calculated for the pristine and for strained systems, as function of
temperature. Substrate effects are discussed for a specific example (SrTiO$_3$).

In particular, we present the structure model in section \ref{model}, and
computational details in section \ref{computational}. The electronic structure and
the thermoelectric properties are addressed in sections \ref{pristine} and \ref{strain}
for pristine and strained CCO, respectively. Section \ref{sto} is devoted to substrate
effects, and some conclusions are given in the final section \ref{summary}.

\section{Structure model}
\label{model}

\begin{figure}[tbp]
\includegraphics[width=0.32\textwidth]{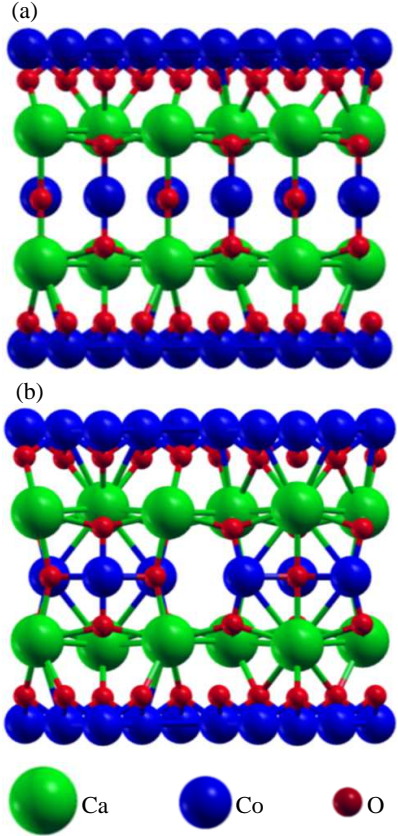}
\caption{Structure (a) before and (b) after the relaxation ($5/3$ ratio of CoO$_2$ and
Ca$_2$CoO$_3$).}
\label{structure}
\end{figure}

As mentioned above,
the monoclinic structure of misfit Ca$_3$Co$_4$O$_9$ is built of rock-salt
Ca$_2$CoO$_3$ sandwiched between hexagonal CoO$_2$ along the $c$ direction. 
The subsystems share the $a$ and $c$ lattice parameters. The $b$ lattice
parameters differ so that a rational approximation with a minimal
lattice mismatch has to be chosen. Asahi et al.\ \cite{Asahi02}
have used $b_1/b_2 = 3/2$ which corresponds to a mismatch of 7\%, whereas
the 5/3 ratio employed by R\'ebola et al.\ \cite{Rebola12} reduces
the mismatch to 3\%. Contradicting experiment \cite{Takeuchi04a,Takeuchi04b}, the 3/2 ratio results in
$3d$ states of the Ca$_2$CoO$_3$ subsystem at the Fermi energy, while the 5/3 ratio leads
only to minor contributions. Thus we consider $b_1/b_2 = 5/3$.

The $a$ and $c$ lattice parameters initially are set to the experimental
values, whereas for the $b$ lattice parameter a value of $5\times2.82$ \AA\ = 14.1 \AA\
(taken from the CoO$_2$ subsystem)
is used. The supercell then is fully optimized, with an on-site Coulomb interaction
of 5 eV. The unrelaxed and relaxed structures are presented
in Fig.~\ref{structure}. The optimized lattice parameters and corresponding bond
lengths are summarized in Table I which also includes a comparison to experimental data.
The optimized $b$ lattice parameter is found to be determined by the CoO$_2$ subsystem
($b_2$), whereas the Ca$_2$CoO$_3$ subsystem ($b_1$) turns out to be intrinsically strained.
Figure 1 indeed shows strong distortions in the rock-salt subsystem.

\begin{table}[t]\centering
\begin{tabular}{c|c|c|c|c|c|c|c|c|c}
 & $a$ & $b_1$ & $b_2$ & $c$ & $\beta$ & Ca-O & Co$_1$-O$_1$ & Co$_1$-O$_2$  & Co$_2$-O (\AA)\\
\hline
GGA & 4.89 \AA& 4.70 \AA& 2.82 \AA &10.73 \AA& 98.4$^{\circ}$&2.46 \AA&2.00 \AA&1.82 \AA&1.92 \AA\\
\hline
GGA+U & 4.88 \AA& 4.71 \AA & 2.82 \AA &10.71 \AA& 98.4$^{\circ}$&2.46 \AA&2.01 \AA&1.81 \AA&1.92 \AA\\
\hline
Exp.\ \cite{Rebola12}& 4.83 \AA& 4.56 \AA & 2.82 \AA&10.84 \AA& 98.1$^{\circ}$&2.40 \AA &2.05 \AA &2.30 \AA&1.92 \AA
\end{tabular}
\caption{Calculated and experimental structure parameters.}
\end{table}

\begin{figure}[tbp]
\includegraphics[width=0.32\textwidth,clip]{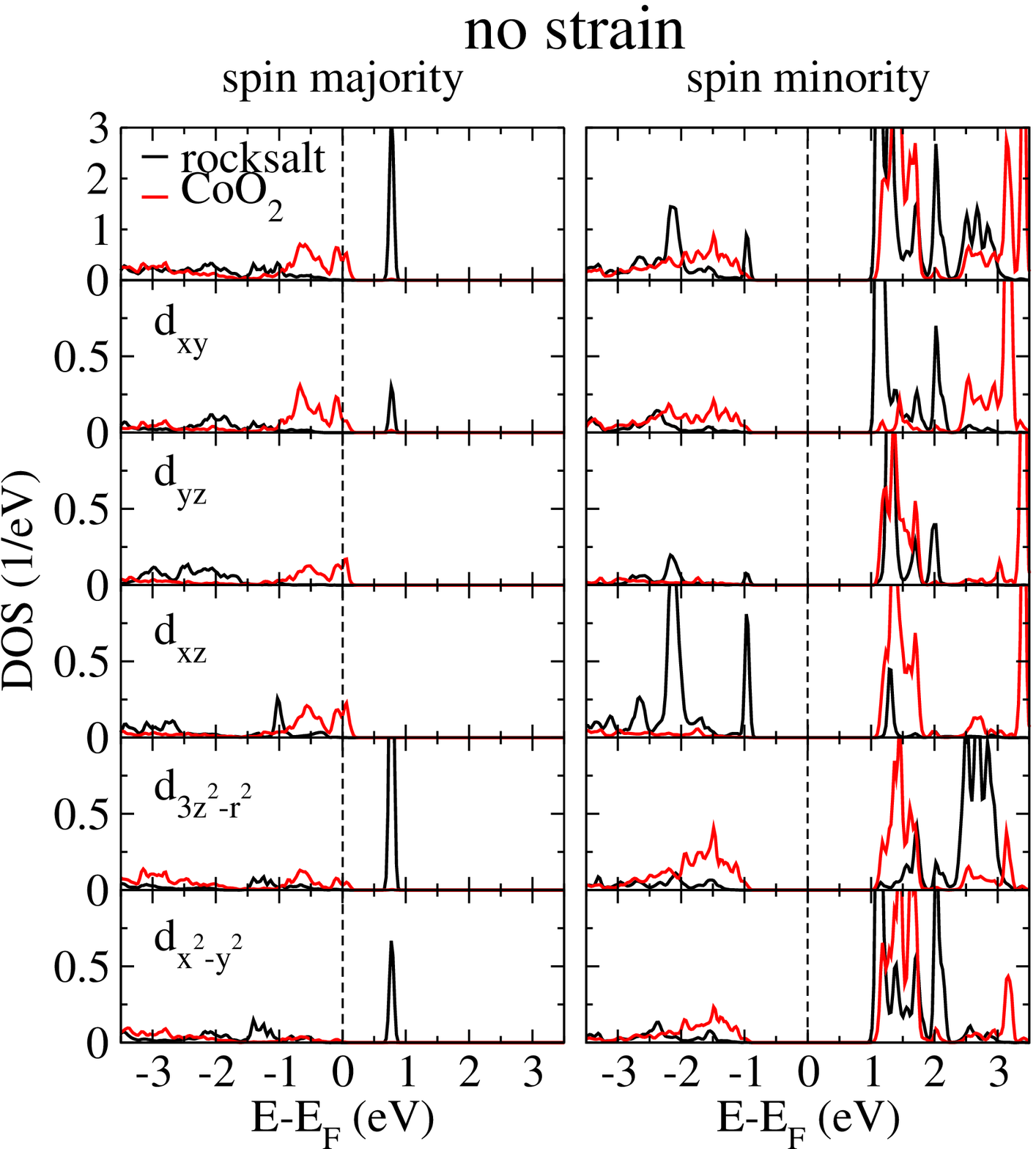}
\includegraphics[width=0.32\textwidth,clip]{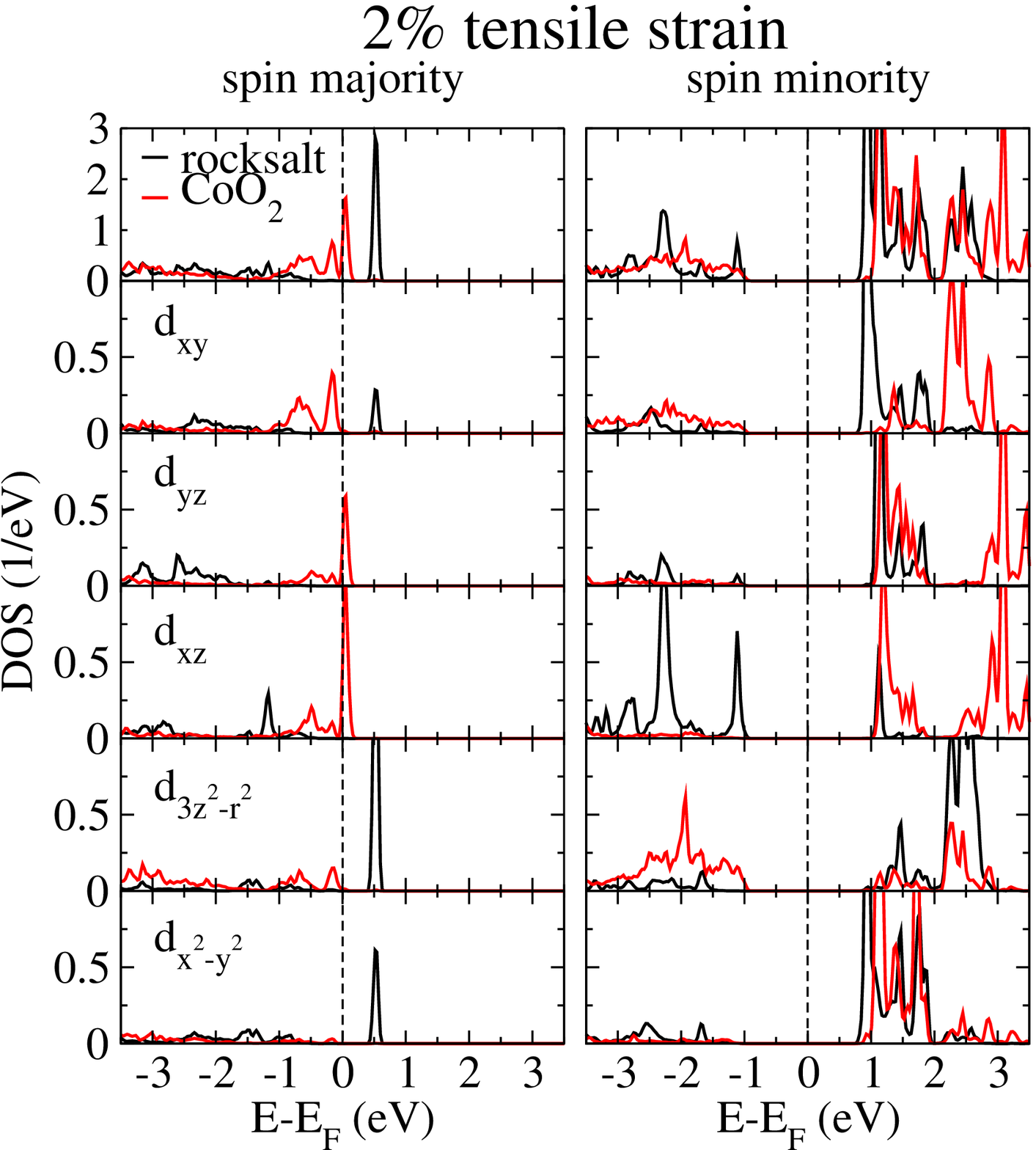}
\includegraphics[width=0.32\textwidth,clip]{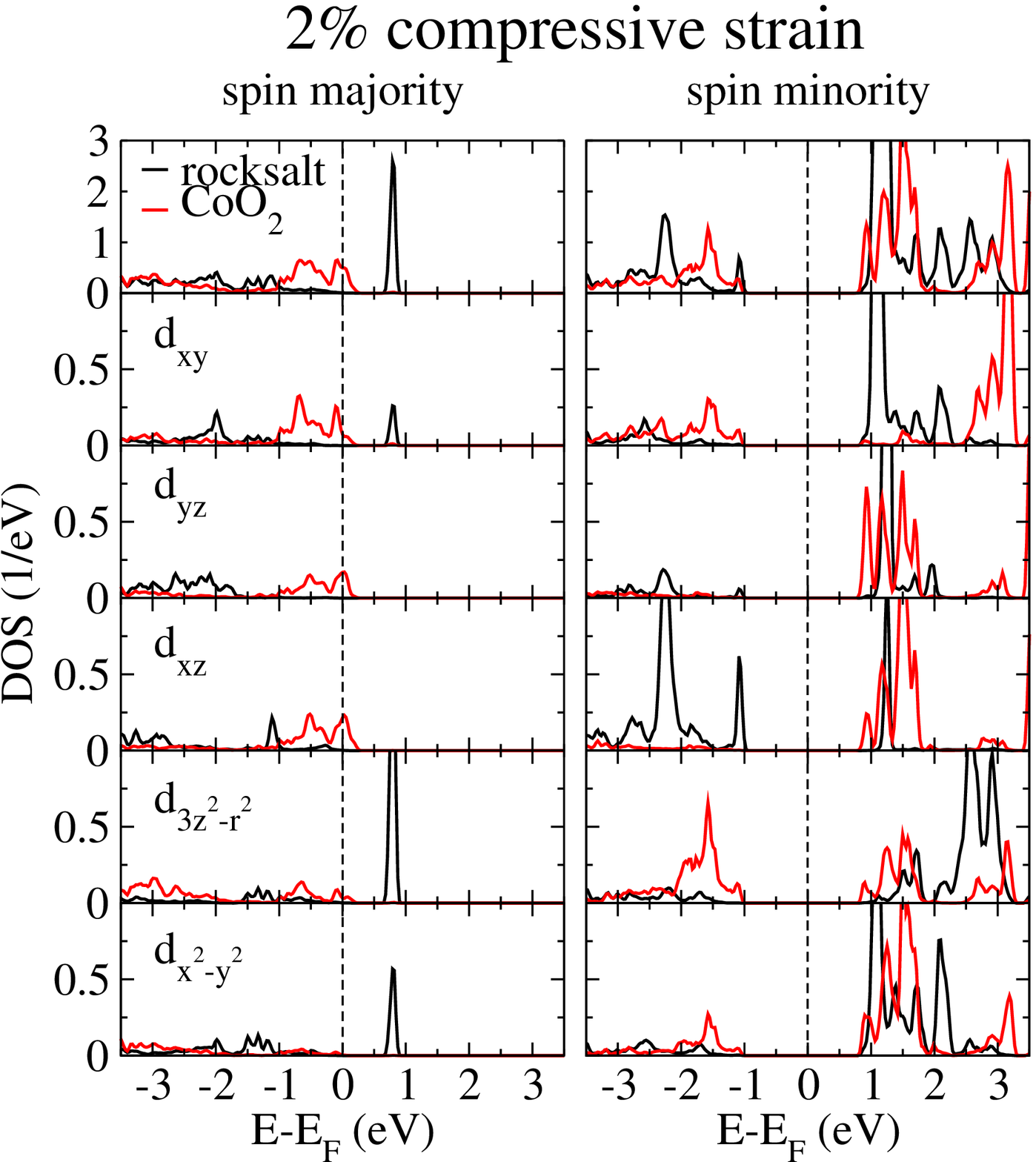}
\caption{Spin majority and minority densities of states without strain (left), under
2\% tensile strain (middle), and under 2\% (right) compressive volumetric strain. The DOS
for 4\% compressive strain is found to be almost indistinguishable from the latter.}
\label{dos}
\end{figure}

\section{Computational details}
\label{computational}
For the structural optimization we use the projector augmented wave method as implemented
in the VASP code \cite{Kresse93}, relaxing both the atomic positions and lattice parameters.
A plane-wave cutoff energy of 530 eV and a residual force criterion of 0.02 eV/\AA\ are used
together with a Monkhorst-Pack $6\times2\times3$ k-mesh. The generalized gradient
approximation in the Perdew-Burke-Ernzerhof flavor \cite{Perdew96} is employed, and spin-polarization
is taken into account.

In order to model the electronic correlations affecting the localized Co $d$ states,
on-site Coulomb repulsions of $U = 3$ eV and 5 eV (with $J = 1$ eV) have been tested.
For $U = 5$ eV the density of states is found to be in good agreement with the
experiment \cite{Takeuchi04a,Takeuchi04b}, whereas for $U = 3$ eV states of the Ca$_2$CoO$_3$ subsystem
appear at the Fermi energy. It turns out that the self-consistent solution
depends critically on the initial value of the Co magnetic moment. Values of
1, 1.5, and 2 $\mu_{B}$ have been compared. In the former (1 $\mu_{B}$) 
case no substantial
magnetic moments are formed in the CoO$_2$ subsystem, and the total energy is strongly
reduced. Thus, only this case is further addressed. Both 2\% and 4\% 
compressive and 2\% tensile volumetric strain are studied, for which the atomic positions
are re-relaxed, respectively.

Precise electronic structures are obtained by the Wien2k \cite{Blaha01} code using
$U=5$ eV and $J=1$ eV. The parameter $R_{mt}K_{max}$ is set to 8, with $G_{max}=24$,
and 320 k-points are considered in the irreducible wedge of the Brillouin zone.
The transport properties are calculated using semi-classical Boltzmann theory, as
implemented in the BoltzTraP code \cite{Madsen03,Madsen06}, employing 2000 k-points
in the irreducible wedge of the Brillouin zone.

We expect this procedure of calculating the transport coefficients to be applicable in
the intermediate metallic regime, hence we show results only for $T$ larger than 200 K.
Furthermore, within the present approach, neither the structural change near 400 K \cite{Wu12}
nor the spin state transition above 600 K \cite{Altin14,Karaman16} is covered.

\section{Electronic structure and thermoelectric properties}
\label{pristine}

The density of states in the top row of Fig.~\ref{dos} (left) shows that only the spin up Co $3d$ states
of the CoO$_2$ subsystem contribute at the Fermi energy, in agreement with experiment \cite{Takeuchi04a,Takeuchi04b}. 
According to the orbitally projected densities of states these
contributions are mainly due to the $t_{2g}$ ($d_{xy}$, $d_{xz}$,
and $d_{yz}$) orbitals. Fig.~\ref{dos} (left) qualitatively agrees with the findings of
Ref.~\onlinecite{Rebola12}; minor deviations can be attributed
to the different computational schemes employed.

\begin{figure}[tbp]
\includegraphics[width=0.5\textwidth]{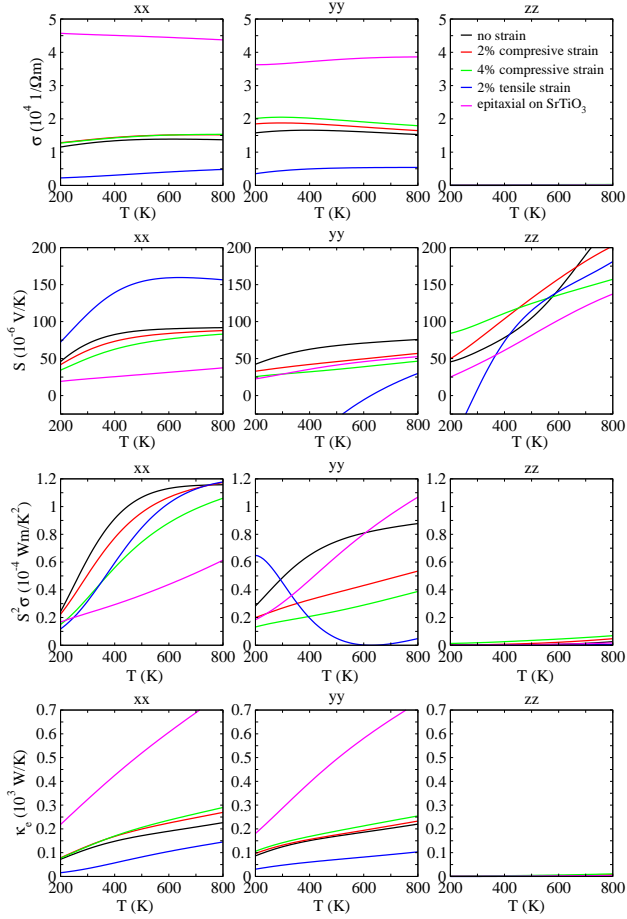}
\caption{Electrical conductivity ($\sigma$), Seebeck coefficient ($S$), power factor($S^2\sigma$), and electronic part
of the thermal conductivity ($\kappa$).}
\label{transport}
\end{figure}

The thermoelectric properties are addressed in Fig.~\ref{transport} where the left, middle, and right
columns refer to the $xx$, $yy$, and $zz$ components, respectively. Comparison of the calculated
electrical conductivity at 300 K with the experiment in Ref.~\onlinecite{Xu02}
($\rho \simeq 10.7$ $\mathrm{m}\Omega\cdot\mathrm{cm}$, where $\rho$ denotes the resistivity)
leads to a relaxation time of $3\cdot10^{-16}$ s,
which is used in the following. We note that the conductivity is enhanced in the 
$a$-$b$ plane (average of the $xx$ and $yy$ components). A maximal conductivity is achieved 
around 400 K. Positive values of the Seebeck coefficient reflect a hole character of the majority
charge carriers. In the $a$-$b$ plane the Seebeck coefficient increases with temperature and 
starts saturating around 400 K, which agrees well with the experiment \cite{Masset00}, whereas 
the $zz$ component does not saturate. At 300 K we find for the total Seebeck coefficient 
(average of the $xx$, $yy$, and $zz$ components) a value close to 60 {$\mu$V}/K, whereas in 
Refs.\ \onlinecite{Asahi02} and \onlinecite{Rebola12} values of 47 {$\mu$V}/K and 227 {$\mu$V}/{K},
respectively, were predicted; and measurements result in 135 {$\mu$V}/K \cite{Masset00,Hu05,Qiao11}. 
The power factor largely resembles the temperature dependence of the Seebeck coefficient.

\section{Strained CCO}
\label{strain}

The densities of states obtained for strained CCO are shown in Fig.~\ref{dos}, for 
2\% tensile strain in the middle part, and 2\% compressive volumetric strain in the right part of 
the figure, respectively. In both cases, similar to the unstrained system, only the spin up Co $3d$ states 
of the CoO$_2$ subsystem contribute at the Fermi energy, while the Ca$_2$CoO$_3$ subsystem gives 
essentially no contributions at all. The same applies to the case of 4\% compressive volumetric strain 
(not shown). The strained systems reproduce the gross features of the unstrained system, except for a 
shift of the spin down conduction band towards the Fermi energy for increasing compressive strain.

The electrical conductivity increases slightly with increasing compressive strain, see Fig.~\ref{transport}.
The overall behaviors of the Seebeck coefficient and power factor as functions of the
temperature do not change under compressive strain, but the values clearly decrease,
particularly for the $a$-$b$ plane. Interestingly, under tensile strain the electrical
conductivity decreases significantly (because of an energetic shift and enhanced
localization of the Co $3d$ states of the CoO$_2$ subsystem at the Fermi energy,
see Fig.~\ref{dos} (middle)), while the in-plane Seebeck coefficient increases at high temperature. 
The total Seebeck coefficient increases at high temperature more strongly than under 
compressive strain but less than for the unstrained case. However, the decrease of the electrical conductivity 
leads to a decrease of the power factor under tensile strain. A strong anisotropy in the Seebeck
coefficient in the $a$-$b$ plane is apparent. The overall behavior of the thermal conductivity as
function of strain is similar to the conductivity, except that $\kappa$ increases approximately
linearly in $T$.

\section{Substrate effects}
\label{sto}

It has been demonstrated for various substrates that first a rocksalt Ca$_2$CoO$_3$ buffer
layer of up to 25 nm thickness is formed before the growth of Ca$_3$Co$_4$O$_9$ sets in \cite{Qiao11}.
The effect is stronger on (001)-oriented perovskite substrates than on hexagonal sapphire
substrates, where the buffer layer extends only over few nm. Next to the buffer layer still
many CoO$_2$ stacking faults are observed, which are assumed to enhance the phonon scattering
and thus the Seebeck coefficient in films up to 50 nm thickness \cite{Qiao11}. Due to the presence of the
buffer layer, it seems unlikely that the substrate has a strong effect on Ca$_3$Co$_4$O$_9$.
This point of view is also supported by the lattice parameters obtained from
high-resolution electron microscopy \cite{Hu05}. However, in order to find out if in-plane
strain from the substrate could have any positive effect on the thermoelectric
performance, we study in the following a representative case, SrTiO$_3$, explicitly.

\begin{figure}[tbp]
\includegraphics[width=0.32\textwidth]{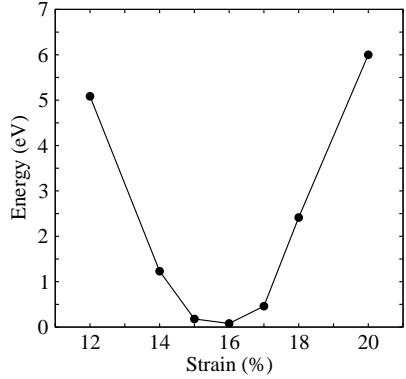}
\caption{Optimization of the $c$ lattice parameter assuming
20\% and 17\% compression, respectively, for the $a$ and $b$ lattice parameters.}
\label{optimum}
\end{figure}

\begin{figure}[tbp]
\includegraphics[width=0.32\textwidth]{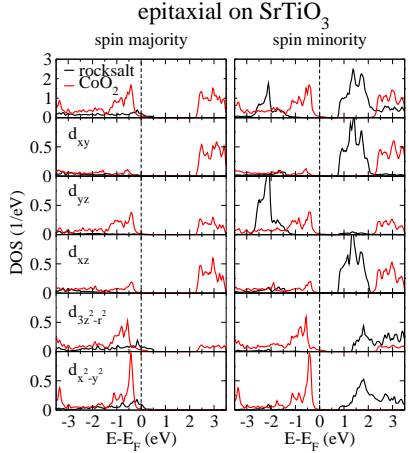}
\caption{Spin majority and minority densities of states when matching the in-plane
lattice parameters to SrTiO$_3$.}
\label{dos-sto}
\end{figure}

To this aim, we strain the Ca$_3$Co$_4$O$_9$ supercell by forcing both the $a$ and $b$ lattice
parameters (original values $a=4.89$ \AA\ and $b=14.1$ \AA) to the lattice parameter of
SrTiO$_3$ (3.905 \AA\ and $3\times3.905$ \AA). This corresponds to rather strong compressions 
of 20\% for $a$ and 17\% for $b$.
We then optimize the $c$ lattice parameter relaxing the atomic positions, see Fig.~\ref{optimum}, 
and obtain an expansion of 16\% to 12.45 \AA.
Due to the strong distortions the densities of states in Fig.~\ref{dos-sto} show significant differences
compared to Fig.~\ref{dos} (right). The consequences for the thermoelectric properties are demonstrated 
in Fig.~\ref{transport} by
pink lines. As expected from the strong compression, the electrical and thermal conductivities
increase, whereas the Seebeck coefficient and the power factor show no improvement over the
unstrained material for most temperatures. These results indicate that Ca$_3$Co$_4$O$_9$
samples are not intrinsically strained and that such strain even would not be fruitful. Though
it is computationally very demanding, a simulation of stacking faults could give further
insights.

\section{Conclusion}
\label{summary}

In this work we have determined, on the basis of a DFT+$U$ calculation which predicts a metallic ground state
for the misfit cobaltate Ca$_3$Co$_4$O$_9$, the temperature dependence of the thermoelectric transport coefficients 
in the constant-relaxation-time approximation. We studied the pristine system (i.e., the 5/3 rational approximation)
as well as its behavior under tensile and compressive strain. The results are expected to be applicable for not too
low temperature, $T \gtrsim 200$ K. In order to explain the measured room-temperature resistivity \cite{Xu02}, we 
have chosen the relaxation time to be given by $3\cdot10^{-16}$ s, slightly smaller than the value used in
Ref.\ \onlinecite{Lemal17}, $8\cdot10^{-16}$ s. Expressed in a different way, we find $\sigma/\tau$ to be given
by $\simeq 0.37\cdot 10^{15}$ $(\mathrm{m}\Omega\cdot\mathrm{cm}\cdot\mathrm{s})^{-1}$, which is by about a factor of
four larger than the corresponding value of Lemal et al. \cite{Lemal17}. Note, however, that the reported resistivities
vary considerably, ranging from $\simeq 2$ for single crystals \cite{Shikano03} to $\simeq 170$ $\mathrm{m}\Omega\cdot\mathrm{cm}$
for thin films fabricated on an LAO substrate using the PLD technique \cite{Jood13}. Another group, also using
PLD but on (001)-silicon with a thin epitaxial yttria-stabilized zirconia (YSZ) buffer layer, reports 
$\simeq 2 - 13$ $\mathrm{m}\Omega\cdot\mathrm{cm}$, depending on the microstructure \cite{Kraus14}. 
Clearly these results indicate a strong dependence of transport properties on the growth conditions and the choice of 
substrate. (Note that we discuss in this paragraph only in-plane transport at room temperature.)

Finally we note that the conductivities found in CCO are considerably smaller than what is observed for other layered
compounds, e.g., the delafossites PdCoO$_2$ and PtCoO$_2$ \cite{Ong10,Gruner15}, which might raise questions
concerning the applicability of Boltzmann transport theory and/or the constant-relaxation-time approximation. In this
context we also recall the simple fact that $\tau \lesssim 10^{-15}$ results in $\hbar/\tau \gtrsim 0.7$ $\mathrm{eV}$
which, naively speaking, would imply a considerable linewidth, comparable to the relevant energy scales in the
density of states discussed above, in the electronic Green's functions. In our opinion, this clearly indicates
that not all aspects of transport in CCO are sufficiently well understood. Luckily, the relaxation time cancels in
the expression of the thermopower---but, even on the Boltzmann equation level, only in the simplest approximation.

\acknowledgments{We acknowledge helpful discussions with Helmut Karl, Anke Weidenkaff, and Karol I.\ Wysoki\'nski, 
as well as financial support by the Deutsche Forschungsgemeinschaft (through TRR 80). 
Research reported in this publication was supported by the King Abdullah University of Science and Technology (KAUST).}

\end{document}